%% file: main.tex
\documentclass[runningheads]{llncs}
\usepackage[T1]{fontenc}
\usepackage{textcomp}
\usepackage{array}
\usepackage{caption}
\usepackage{verbatim}
\usepackage{graphicx}
\graphicspath{{figures/}}
\usepackage{booktabs}
\usepackage{multirow}
\usepackage[hyphens]{url}
\usepackage{hyperref}
\usepackage{xcolor}
\usepackage{makecell}
\usepackage{enumitem}
\usepackage{listings}
\newlist{tabitem}{itemize}{1}
\setlist[tabitem]{wide=0pt, nosep, leftmargin= * ,label=\textbullet,after=\vspace{-\baselineskip},before=\vspace{-0.6\baselineskip}}

\usepackage{pifont}

\usepackage{csquotes}

\def\eg{e.\@g.\@,~}
\def\ie{i.\@e.\@,~}
\def\etal{et~al.\@}

\newcommand\quoting[2]{``\emph{#2}'' (#1)}
\begin{document}

%
\title{Privacy in ERP Systems: Behavioral Models of Developers and Consultants\thanks{This paper was presented at the 20th International Conference on Risks and Security of Internet and Systems (CRISIS'2025) and will appear in its post-proceedings. This is the author's version.}}

\titlerunning{Privacy in ERP Systems}
%
\author{Alicia Pang\inst{1} \and
Katsiaryna Labunets\inst{2} \and
Olga Gadyatskaya\inst{1}}
\authorrunning{A. Pang et al.}

\institute{Leiden University, the Netherlands \\
\and
Utrecht University, the Netherlands \\
%
}

\pagestyle{plain}
\maketitle

\begin{abstract}
Applications like Enterprise Resource Planning (ERP) systems have become an indispensable part of the corporate digital infrastructure. These systems store sensitive data about customers, suppliers, and employees, and thus companies have to process these data in accordance with applicable regulations like the GDPR (the EU General Data Protection Regulation). This can be challenging due to a variety of reasons. For example, prior research has shown that developers sometimes lack knowledge about privacy. 

In this work, we focus on privacy in ERP systems in the context of an international consultancy firm. We investigate the privacy awareness regarding privacy-by-design and data minimization of two important populations: developers of ERP systems and managers and consultants responsible for services related to ERP systems. Applying thematic analysis, we elicit privacy behavioral models of these two populations using Fogg's Behavioral Model (FBM) framework. Our findings provide a means to stimulate more adequate privacy-related behaviors for developers and consultants. 
\end{abstract}


\input{sections/introduction}
\input{sections/background}

\input{sections/methodology}
\input{sections/results}

\input{sections/model}

\input{sections/discussion}

\input{sections/related_work}

\input{sections/conclusions}

\subsection*{Acknowledgments}
This research has been partially supported by the Dutch Research Council (NWO) under the project NWA.1215.18.008 Cyber Security by Integrated Design (C-SIDe).

\bibliographystyle{splncs04}
\bibliography{bib}

\end{document}

%% file: sections/introduction.tex
\section{Introduction}

Due to fast-paced technology developments, companies around the world collect and share large amounts of personal data. However, technological developments also provide new opportunities for cyberattacks, such as identity theft or data breaches~\cite{verizon2025}. As a result, protecting the privacy of personal data has become extremely important to protect individuals~\cite{gdpr2016EU}. To better protect personal data, in 2018 the EU introduced the General Data Protection Regulation (GDPR), which is applicable to the personal data of all EU residents, even if the processing company is based overseas. However, not all companies are aware of the importance of data protection and the challenges that the GDPR brings~\cite{dalela2021mixed,birrell2024sok}.

The GDPR is underpinned by several important principles such as \emph{data minimization} and \emph{privacy-by-design} (PbD). Integrating privacy into the structure of the organization is one way to change people's behavior. As a result, PbD is treated as a core strategy for the entire organization~\cite{cavoukian2009privacy}. However, there are still some challenges to overcome when implementing it: this concept has been characterized as ``vague'', leaving many unanswered concerns relating to how to apply it in system design~\cite{vanrest2014designing,spiekermann2012challenges}. Researchers and engineers tend to associate the PbD concept with certain privacy-enhancing technologies (PETS)~\cite{hadar2018privacy}. On the other hand, privacy-by-design cannot be reduced to a set of rules or the use of a specific technology, as it is a process that involves a variety of technological and organizational measures that enforce privacy and data protection principles through the use of appropriate and adequate technical and organizational methods, including PETS~\cite{castelluccia2022data}.

It has emerged that key stakeholders in organizations like developers and managers do not yet have a clear
understanding of privacy, implications of GDPR, or what the concepts of privacy-by-design and data
minimization mean for their work~\cite{birrell2024sok,hadar2018privacy,iwaya2023privacy,vanrest2014designing,vanderlinden2021data}. In our research, we study \emph{perceptions and behaviors regarding these important concepts and the GDPR in general} of two populations: \emph{developers} who develop and maintain ERP systems (which includes creating specific functionalities based on customer
requirements and solving technological problems) and \emph{consultants and managers}  who actually
support the customers and help to set up ERP systems. In addition, managers and consultants are responsible for privacy protection and ensuring the confidentiality of personal data\footnote{In this study, the only difference between a manager and a consultant is that a manager can manage a consulting project independently, whereas a consultant is supported by a (senior) manager. Thus, we group people in these roles.}. This study was done at a multi-national consultancy company in the Netherlands. We focus on ERP systems because this is an important type of system ubiquitously used by enterprises to handle sensitive data.

We performed a qualitative study and collected data via 16 semi-structured interviews with the target populations. We analyzed the data using thematic analysis and we mapped the identified concepts and themes to BJ Fogg's Behavioral Model~\cite{fogg2009behavior}. In this way, we discovered the key perceptions, motivations, ability factors and triggers that make it (im)possible for our study groups to perform data minimization, PbD, and other privacy-related tasks in their jobs. Our study sheds light on the factors that affect the implementation of privacy protection for ERP systems. 

%% file: sections/background.tex
\section{Background}\label{sec:background}

\paragraph{GDPR and related principles.}
The General Data Protection Regulation (GDPR), also known as Regulation (EU) 2016/679, is a privacy law introduced by the European Parliament that aims to give European residents more control over their personal information~\cite{gdpr2016EU}. 
To protect personal data, the GDPR establishes seven principles that must be followed when collecting, processing, or controlling personal data. These principles are \emph{lawfulness, fairness and transparency}; \emph{collection for a specified and legitimate purpose}; \emph{data minimization} (collected data is relevant, adequate and limited); \emph{accuracy}; \emph{limited retention}; \emph{secure processing}; and \emph{accountability}~\cite{gdpr2016EU}. 
Software systems, such as ERP systems, process personal data and therefore must adhere to these principles to comply with the GDPR. 

The enactment of the GDPR created many challenges for organizations, as the existing data processing procedures had to be modified or re-developed, with new roles and tasks introduced into existing business processes. Additionally, the GDPR specifies the principles but does not mandate how they should be implemented technically, leading to doubts and loopholes~\cite{kutylowski2020gdpr}. For example, Art. 25 of GDPR stipulates that organizations must follow \emph{privacy-by-design} (PbD) and \emph{privacy-by-default} paradigms when processing personal data -- but it refrains from suggesting concrete technical implementation measures, except pseudonymization, which is provided as an example. 

Given the complexity of GDPR, in this research, we focus only on the two core GDPR concepts -- \emph{data minimization} and \emph{privacy-by-design} -- and we investigate how these principles are understood and operationalized by professionals (developers and consultants) in a large service company.


\emph{Data minimization} (DM) is a principle that advises minimizing the use of personal data in software systems~\cite{senarath2018understanding}. Although it appears to be straightforward, the DM principle can present a challenge, as, for example, it can be in conflict with the business needs, as, for example, collecting as much customer data as possible can enable better marketing strategies~\cite{andrew2021general}. 

\emph{Privacy-by-design (PbD)} refers to a technical and strategic management approach that commits to choosing and implementing governance controls to reduce the privacy risks of information systems~\cite{spiekermann2012challenges} and implies that privacy must be considered throughout the whole design process~\cite{cavoukian2009privacy}.

\paragraph{Enterprise Resource Planning (ERP).} 
ERP systems are business information systems designed to manage all resources, information, and relevant tasks for the entire business operations. Nowadays, even small and medium-sized businesses use ERPs~\cite{hasan2018impact}. These systems are available from different vendors, with Microsoft, SAP, and Oracle being the most popular. Common ERP systems integrate modules such as production planning, purchasing, supply chain, inventory management, human resources, accounting, marketing, and finance. 

A software solution such as an ERP system that collects, saves, and analyzes personal data (e.g., customer data) obviously needs to comply with data protection laws, including the GDPR.
Personal data is often not marked explicitly as such, making it difficult to find it among hundreds of tables in the ERP system. It can take a lot of time and effort to find and manage personal data in systems of this scale and complexity~\cite{arachchi2015quality}.

\paragraph{Behavior models.}

Theories of behavior change and behavioral models have been put forward to help develop interventions to promote good habits and minimize harmful behaviors. In an attempt to explain behavior change, each behavioral change theory or model focuses on different aspects. The social cognitive theory, theories of reasoned action, the trans-theoretical model of behavior change, the health action process model, the COM-B model~\cite{michie2011behaviour}, and BJ Fogg's Behavior Model (FBM)~\cite{fogg2009behavior} are among the most common frameworks~\cite{taj2019digital}. 

In this study, we use FBM to investigate the privacy behaviors of developers and consultants.
According to FBM~\cite{fogg2009behavior}, human behavior is the result of three factors: \emph{motivation}, \emph{ability}, and \emph{trigger}. In short, behavior occurs when someone wants to do something (motivation), is able to do it easily (ability), and there is something that drives the action (trigger or prompt)~\cite{fogg2009behavior}.

%% file: sections/methodology.tex
\section{Research Approach}\label{sec:methodology}

In this study, we collected data via semi-structured interviews with the target populations. 
All interviews were conducted online via Microsoft Teams; the average duration of an interview was about 1 hour. The interviews were recorded and then transcribed. Some interviews (when both parties were native Dutch speakers) were conducted in Dutch; transcripts of these interviews were then translated into English for analysis.

Our interview protocol contained 3 segments. In the first segment, demographic information of the interviewee was collected, and their suitability as a participant was confirmed. In the second segment, the interviewees were asked questions related to their understanding of privacy, DM, and PbD, and their responsibilities related to data protection. In the final segment, we asked questions focused on the privacy challenges for ERP systems, the participants' experience, and responsibilities related to implementing DM and PbD in ERP systems, and what challenges they had in that respect.

\paragraph{Participants.}
This research was conducted within an EU office of a multi-national consulting company that provides, among others, consulting services on setting up and managing ERP systems to various organizations.

\textbf{Recruitment.} We used convenience sampling to recruit study participants at the consultancy firm. 
We sent an e-mail to a potential interviewee in a relevant role, informing them about the study and asking when they would be available
for an interview. An invitation to the interview and a consent form were
sent to the respondents who replied positively. The consent form contained further information
regarding the study and asked for the respondent's consent to participate and for us to record the interview for transcription purposes. No remuneration was provided for participation. 

\textbf{Demographics.}
We recruited and interviewed
9 consultants/managers and 7 developers (16 people in total). We checked with them that they qualified as participants in our study based on their position (role), experience in working with ERP systems, involvement in privacy-related decisions, and level of education. 
The demographic information of the participants is summarized in~\autoref{tab:demographics}.
In the remainder of the paper, participants in the developer role have IDs starting with \textit{D} (e.g., D1), consultants have IDs starting with \textit{C}, and managers have IDs starting with \textit{M}.

\begin{table}[t!]
\caption{Summary of participants' demographics.}
\label{tab:demographics}
\scriptsize
\centering
\begin{tabular}{p{1cm}p{1cm}p{1cm}|p{3cm}p{1cm}p{1cm}|p{1cm}p{1cm}p{1cm}}
\toprule
\multicolumn{3}{c|}{\textbf{Level of education}} &
\multicolumn{3}{c|}{\textbf{Role}} &
\multicolumn{3}{c}{\textbf{Work experience}} \\
\midrule
 & \textbf{C/M} ($n$=9) & \textbf{D} ($n$=7) 
 & & \textbf{C/M} ($n$=9) & \textbf{D} ($n$=7)
 & & \textbf{C/M} ($n$=9) & \textbf{D} ($n$=7) \\
\midrule
BSc & 3 (33\%) & 4 (57\%) 
& ERP developer & -- & 3 (43\%) 
& Min & 2 years & 2 years \\
MSc & 6 (67\%) & 3 (43\%) 
& Software architect & -- & 2 (29\%) 
& Median & 10 years & 9 years \\
& & 
& Data engineer & -- & 1 (14\%) 
& Max & 31 year & 14 years \\
& & 
& Internal advisor & 1 (11\%) & -- 
& & & \\
& & 
& Legal council & 1 (11\%) & -- 
& & & \\
& & 
& (Senior) consultant & 2 (22\%) & -- 
& & & \\
& & 
& (Senior) manager & 5 (56\%) & 1 (14\%) 
& & & \\
\bottomrule
\multicolumn{9}{l}{Note: C/M - Consultant/Manager; D - Developer} \\
\end{tabular}
\end{table}

\paragraph{Data analysis.}
To analyze the data for this research, three researchers first conducted a five-phase thematic analysis following Braun and Clarke~\cite{braun2006using}. In the first phase, \emph{familiarization with the data}, three researchers read the transcripts to understand their content. In the second phase, \emph{generation of initial codes}, one coder systematically analyzed the transcripts and inductively coded them, forming the initial codebook, which was reviewed by two independent researchers. In the third phase, \emph{searching for themes}, the three researchers identified recurring concepts and patterns in the data, grouped codes, and named the found themes. In the fourth phase \emph{reviewing themes}, the themes were examined for being coherent and relevant, and refined. In the fifth phase \emph{defining and naming themes}, the specifics of each theme were further refined, and themes were clearly defined. During these phases, the researchers met frequently to discuss the codebook and the themes, and all disagreements were resolved. To ensure the quality and reliability of the final codebook, one of the researchers (not the original coder) used the final codebook to code two randomly selected transcripts. This resulted in Krippendorff's $\alpha$ equal to 0.971, indicating a very high agreement~\cite{atlasti2020}. 

Once the codebook was finalized and the themes were defined, these were then analyzed by the three researchers together and deductively labeled into \emph{motivations}, \emph{ability}, and \emph{triggers} following FBM~\cite{fogg2009behavior}, as explained in Sec.~\ref{behaviour-model-1}. We used the \texttt{Atlas.ti} software for coding the data (without the AI features).

\paragraph{Ethical considerations.} The study design was approved by the Science Ethics Committee at Leiden University. All participants were informed about the goals of the study and the data we collected, and consented to participation prior to the interview. Participants were not remunerated. 

\paragraph{Limitations.}
There are limitations to this study that need to be discussed. First, our sample is limited, as we only interviewed employees from a single consultancy service company in a single EU country. Culture and business practices within the company and the national regulatory and cultural landscape may affect our findings. Furthermore, as we used convenience sampling, our participants were motivated to participate, and they might be more privacy-aware or consider it more important than other employees. Future research replicating this study across various organizations and countries could address this limitation. Social desirability bias is another potential limitation of this study. We rely on self-reported behaviors and perceptions of our respondents, and they might have been provoked by the interview setting to inflate the importance of privacy and security in their job. We tried to calibrate for that by also asking questions about factual knowledge (e.g., their own definitions of data minimization and privacy by design) and asking for clarifications and examples, where possible. 

%% file: sections/results.tex
\section{Results}\label{results}

\textbf{The role of GDPR in the work of developers and consultants:} For some of our respondents, the role of GDPR is quite significant. Some developers stated that they had to develop extra functionalities in the ERP system to ensure the application was GDPR-compliant. Another developer stated that they and their colleagues take GDPR very seriously and must abide by it:
\quoting{D4}{If one of our developers doesn't abide by GDPR, they might end up losing their job. Or there will be strict actions taken against them. So the data protection is very key. And we all, as a developer, as a consultant, we abide by it, and we have to follow it}. GDPR was also a major driver for privacy management in ERP systems: \quoting{D4}{I have never bothered about privacy [before GDPR]. Because nobody told me that it was important in ERP. I didn't know this}. 

However, there is an alternative opinion that GDPR does not directly impact developers' work, but there is more awareness about protecting data confidentiality in general. 
Similarly, managers and consultants noted that the GDPR influenced their work by \emph{increasing privacy awareness} to some extent:
\quoting{C8}{I will not say that there is a very strict follow-up, and the awareness is not quite there yet. When this law was not yet in place, employee data and suchlike were handled more senselessly. But now it's more sensitive}.

Despite the fact that GDPR is mandatory and that it has created awareness, GDPR is not high on the priority
list, and doing a training on it does not sound exciting, as one developer put it: 
\quoting{D2}{I have other business to do and no time for that. The GDPR will be the lowest priority on my list. It's not the most exciting thing either}.

Developers and consultants working with major vendors and existing ERP platforms noticed that many of the suppliers showcase the compliance and trustworthiness of their products with  privacy and data protection certifications:
\quoting{M4}{Microsoft themselves give the system with a full privacy and data protection certificate. So that comes with the tool}.  
As a result, developers and managers do not need to worry much about GDPR compliance in their ERP systems. They consider it to be GDPR-compliant by default because privacy comes ``\emph{out of the box}'' in the available ERP frameworks:
\quoting{M1}{The ERP system is already compliant. So you don't have to be mindful of the GDPR part}.

\textbf{ERP privacy challenges:}
Our participants shared several privacy/GDPR-related challenges for the ERP systems. Firstly, access control is perceived as a data management challenge, but this task is a shared responsibility for all ERP system stakeholders, as developer D1 stated: \emph{``the challenge for us is to classify those roles so they don't see more information than they need. That's an exercise everybody does.''} However, this participant perceives it as imposed by GDPR: \emph{``GDPR is the reason that we have the challenge. Nobody would have been bothered about that.''}

One of the managers stated that one of their responsibilities is to manage access data rights for specific users: 
\quoting{M2}{When I design an authorization, certain people are allowed to see specific data. But also ensure that this data is not visible to everyone}.

Yet, ultimately it is the customer, not the consultant, who decides which roles have access to which data. The consultant advises and supports the customer in the implementation and technical aspects. However, it is the customer who uses the system and is ultimately responsible for what they put into the system:
\quoting{C8}{The use of the system is by the client and also their responsibility with what they put in or not }.

However, our participants observed a number of malpractices among their customers regarding data handling. 
One of the consultants received an email with an Excel file containing all the employees' salaries from their customer. 
Another developer received a database from the customer with all the (personal) data. The developer asked the customer for a test database, but this does not always happen in practice: 
\quoting{D3}{During development, we get files, for example, the payroll interface. So we can see all the wages of companies for that month. Then I always ask for a test file, but the customer does not provide it in practice. They fail, and we get an actual file with that data. It's not OK}.

Managers emphasized the importance of data privacy frameworks and organizational privacy policies, which are designed to protect privacy and minimize data-related malpractices: \emph{``it becomes very important for me to follow the guidelines and principles and work on such projects''} (M1). 
Another challenge that most consultants or managers experience is the the system integration when personal data from one system gets consumed by another system.

\textbf{Data minimization (DM):}
The data minimization concept is relatively familiar to developers: most of our developer participants provided similar definitions of it: \emph{``less mandatory fields are possible [\dots] if you don't need that, you don't need that''} (D1).
Similarly, more than half of the consultants/managers were familiar with the \emph{data minimization} concept. However, some participants did not know or were unsure of the exact meaning of DM, as it played a minimal role in their function.

\textbf{Privacy-by-Design (PbD)} is a somewhat familiar concept for developers and consultants, yet some developers could not explain what PbD means: they had never heard of it and had no (direct) experience with it.
Some consultants and managers were also unaware of what it entails and had no prior experience with PbD. This observation aligns with van Rest \ et al.~\cite{vanrest2014designing}, who noted that PbD remains a vague concept among managers, with its exact meaning often unclear. Meanwhile, some consultants were aware of this term but remarked that they neither work with it nor have any hands-on experience. Despite this, both developers and managers describe the concept of PbD in a similar way: \emph{``In my view, privacy-by-design is already taking privacy into account when implementing or configuring a system''} (M2).

\textbf{Impediments:} Developers and consultants frequently see \emph{privacy as a security problem}. Regarding the privacy goals of the ERP system, the developers were more focused on securing information and restricting access to it than protecting privacy:
\quoting{D2}{The ERP system is mainly used within companies. From the outside
[\dots] you effectively have all the tools to detect if a certain subset of data is sent out. There is also an extensive roles system with security. [\dots] So it's all pretty watertight 
}.

Furthermore, there is often a conflict between business needs and privacy regulations. According to a developer, customer requirements can be challenging, e.g., a client may request specific data, such as date of birth, raising concerns about GDPR compliance. This increases the workload for developers and consultants when designing and configuring ERP systems. 

To conclude, not all developers, consultants, and managers are familiar with privacy concepts like data minimization and PbD. This lack of knowledge makes it challenging to implement these concepts in an ERP system. One key reason for this gap is insufficient training in privacy-related practices: \emph{``No, I haven't seen people [learn] about data minimization much''} (M4).

\textbf{Strategies and techniques for implementing DM and PbD:}
Due to their lack of knowledge, developers and consultants are unfamiliar with the techniques and strategies for implementing PbD and DM in ERP systems, which aligns with the findings by Oetzel \etal~\cite{oetzel2014systematic}. However, privacy protection techniques have already been implemented in ERP systems, as mentioned earlier. As a result, developers and consultants often perceive that they do not have to think about or implement PbD or DM:
\quoting{M1}{But in the ERP system, as I said, these are standard screens that are globally accepted and agreed upon. I don't have to put anything specific to achieve this privacy-by-design}.

Among the mentioned strategies and methods, we can mention
\emph{data retention and deletion policies:} As a developer explained, data is retained in the system for a specified period and then automatically deleted after a certain number of years. The developer’s only responsibility is to define the retention rule, and the system handles the rest.


Moreover, \emph{external support} can play a role: consultants explained that 
whenever a project involves processing a lot of personal data in the ERP system, a data migration team or a dedicated compliance team can be brought in to help. Other common security controls the developers use to safeguard privacy are \emph{multi-factor authentication}, \emph{logs auditing} and \emph{role-based access control}.

Two managers stated that there are \emph{templates/blueprints} available within their consultancy firm. These blueprints explain what the organization thinks about the PbD or DM concepts. Sometimes the blueprints are presented to the customer to check whether this approach is suitable for them.

%% file: sections/model.tex

\section{Behavior Model}\label{behaviour-model-1}

\begin{figure*}[t]
    \centering
    \includegraphics[width=0.72\textwidth]{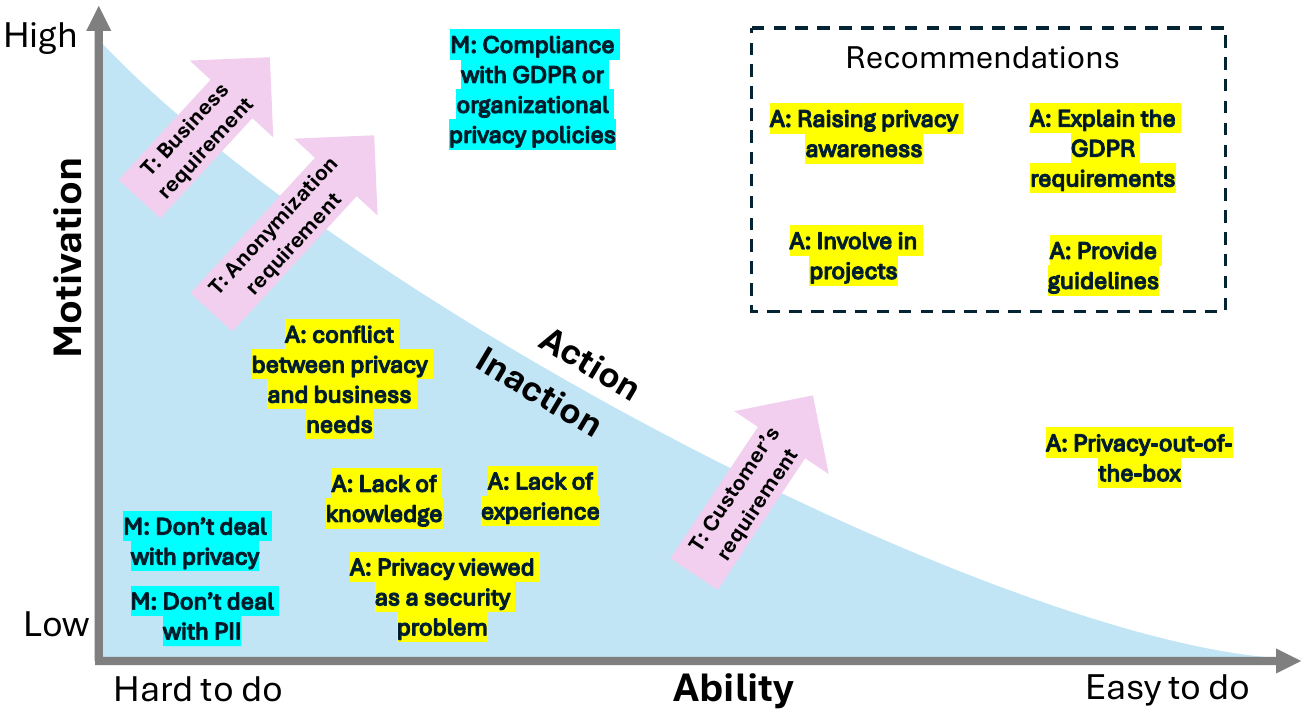}
    \caption{Privacy behavior model for developers. Light blue stands for motivations (M); yellow for ability (A), and pink for triggers (T).}
    \label{fig:developersFBM}

    \vspace{1em} 

    \includegraphics[width=0.72\textwidth]{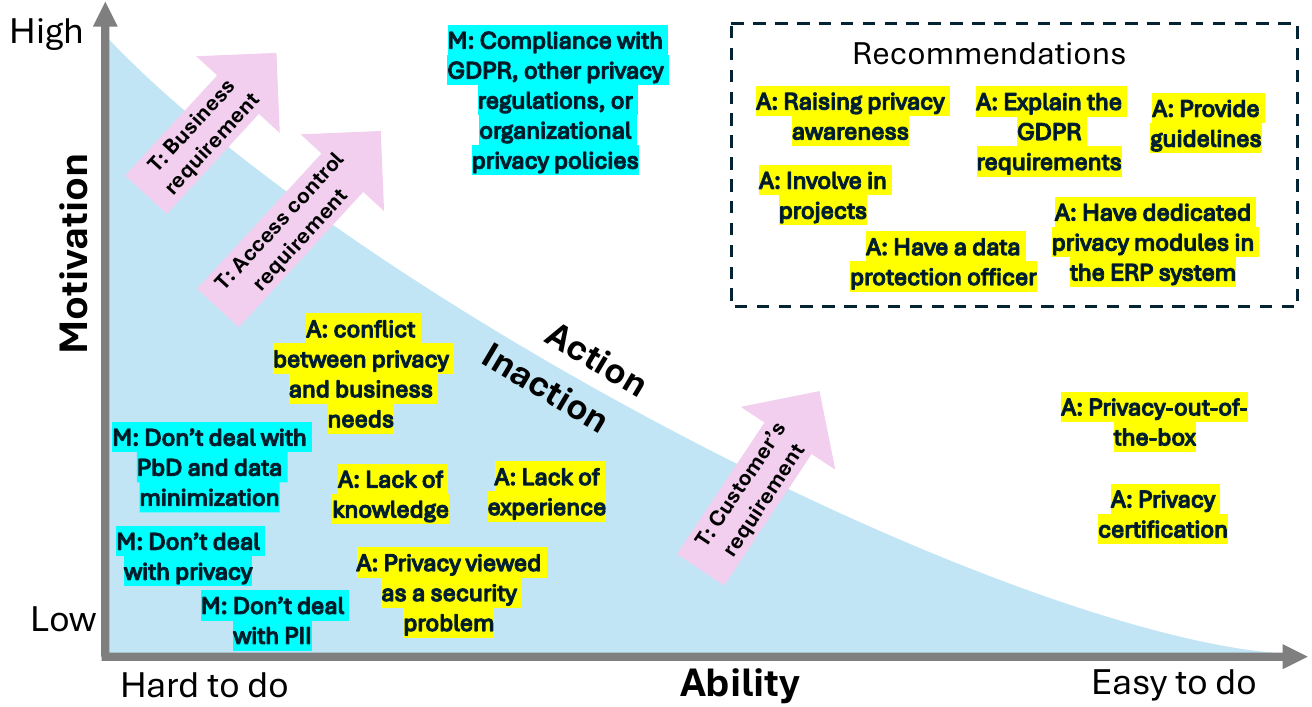}
    \caption{Privacy behavior model of consultants and managers. Light blue stands for motivations (M); yellow for ability (A), and pink for triggers (T).}
    \label{fig:consultantsFBM}
\end{figure*}

Based on the results from the previous section, we now
discuss the behavior of developers, consultants, and managers. The BJ Fogg Model is used
to understand behavior, using the formula: \emph{behavior} = \emph{motivation + ability +
triggers.} We deductively assigned labels related to \emph{motivation}, \emph{ability}, \emph{trigger}, and \emph{general} to the codes in our codebook to identify the aspects of behavior. The resulting behavioral model of developers is illustrated in Figure~\ref{fig:developersFBM}, while Figure~\ref{fig:consultantsFBM} shows the behavioral model of the consultants and managers. For the lack of space, we only discuss some of the most interesting aspects of these models.

\paragraph{Concepts.}
\textbf{Motivation:} Motivation can be described as how much the person is driven to do something or against it. There must be a reason why a specific behavior is being performed. A question that can be asked to gauge whether it is related to motivation is ``am I willing to do this?''~\cite{fogg2009behavior}. 
For example, when an interviewee mentions GDPR compliance or the need to follow some instructions, this motivates them to implement privacy into the system. 

\textbf{Ability:} 
Ability refers to how easy or hard it is for someone to perform something at a particular time. 
We consider that, for example, when an interviewee mentions that they have no knowledge about data minimization, implementing data minimization in the system will be difficult for them. Another example of ability is when an interviewee mentions some system functionality that makes it easier for them to protect privacy.

\textbf{Triggers:}
A trigger is an event that starts the process. There are different triggers that have been described in the literature, such as social, forced, and proactive triggers~\cite{das2019typology}. For example, when an interviewee mentions business or customer requirements, this effectively serves as a trigger to perform a privacy-related task.

\paragraph{Developers' behavior.}

\textbf{GDPR compliance} and \textbf{the organization's privacy policies} are notable motivations for developers to follow the GDPR requirements. If they do not follow the GDPR requirements, there are consequences for their job. Therefore, they are ``\emph{highly}'' driven to follow the GDPR and other policy requirements.

In some ERP systems, little personal information is kept, so developers \textbf{do not deal with or work with personal information in the ERP system.} In that case, the motivation to implement privacy techniques to protect this personal information can be very low. Another (de)motivation can be that the developer (perceives that they) \emph{do not deal with privacy}: their motivation is then low. This is how D2 expresses their thoughts on privacy: \emph{``No, you don't think about privacy at that moment''.}

A \textbf{conflict between privacy and business needs} can affect the
developer's ability. While the business needs something
like creating a function where they can access everything in the system,
this can conflict with privacy laws. This makes the
developer's ability ``\emph{obstructed}''.

\textbf{Mixing different privacy concepts} and a general \textbf{lack of knowledge} can also hinder the
developer's ability. Furthermore, the ability can be obstructed because developers see \textbf{privacy as a security
problem}. Not having the appropriate expertise can make
implementing privacy techniques hard. This is how D1 answered the question whether privacy-by-design is essential
for the ERP system: ``\emph{It's not like public domain data. It is a private network where only people in the organization can see
the data. And with a certain level of clearance.}''

However, \textbf{privacy out of the box} makes the developers' work more accessible. When ERP providers implement privacy
frameworks in the system, developers have less work to do.

The behavior also has to be triggered. An example trigger for a developer can be a \textbf{requirement}. These can be business- or customer-related. \textbf{Business requirements} focus
more on operational needs such as system performance, cost, and storage.
This is how D2 explains that, based on a business requirement,
they need to apply DM:
\quoting{D2}{Not because of privacy but more because of the
database size. The storage is minimized and additional storage also
costs more money for the customer}.

\textbf{Customer requirements} focus more on what the customer demands. Moreover, \textbf{the customer being involved} in projects 
can also trigger the developer's behavior: 
\quoting{D1}{You want to process that data, and also give a
choice to the customers or the people who give you the data to have a
say in how it is managed or how it is stored or how it is
shared}.

\textbf{Improvements of behavior.} 
In FBM, improvement of behavior means increasing the ability or motivation, or adding a trigger, what facilitates moving across the action line. One of the recommendations is to have \textbf{general guidelines}. This may be classified as a motivation or ability for the developer. According to one of the developers, it can be demotivating and annoying if there are no general guidelines or if those are not clear: \quoting{D7}{But having your procedures in place, and simple procedures, not too much complication. When you make things more
complicated, it's just becoming annoying}.

According to one of the developers, getting \textbf{involved in a
project} can increase the ability:
\quoting{D6}{That I'm getting involved in. [...] Everyone is getting a little
bit involved in, 'Hey, we made this and this choice, for
example. [...]' So that you get a detailed
description of how it happens, take people in there and tell them why choices were made}.

\textbf{Raising privacy awareness} by, for instance, following training,
presentations or workshops can change the developer's
ability:
\quoting{D1}{By having training, having a mandatory boot camp,
just to give you the policies of the company. So explain, based on the
rule on what they will have, what they need to do}.

\paragraph{Consultants and managers' behavior.}

Consultants and managers are motivated by \textbf{the GDPR compliance}:
\quoting{C9}{You must
be compliant, and of course, Microsoft must also comply with the
package}. Another motivation for the consultants and managers can be the
\textbf{organization's privacy policy}. For example, one
of the managers explained that it is essential to follow the guidelines
and principles that come from the GDPR. Therefore, it is a ``\emph{high}''
motivation because the manager sees it necessary to follow:
\quoting{M1}{So becomes very important for me to follow the guidelines and principles and work on such projects}.

However, it can also be that the consultants and managers are not
motivated, e.g., if they \textbf{do not deal
with data minimization, privacy-by-design, privacy, or personal
information.} Consultants or managers do not deal with DM
or PbD because it is not part of their responsibility.
This is how one of the managers put this:
\quoting{M2}{I don't have any specific deal with this
data minimization or privacy by design in my day-to-day life}.

The ability of consultants and managers is also impeded by
\textbf{conflicting interests between privacy and business needs.} For
example, M1 explains that the business wants to test with real-life
data, which is not necessarily allowed under the GDPR because it contains personal
information. This obstacle for the manager results in low ability
(``\emph{hard to do}'').

\textbf{The lack of knowledge and the mixing of different privacy
concepts} also obstruct the ability of the managers and consultants to
implement privacy techniques in the ERP system. Moreover, the ability to implement privacy techniques in the ERP system can also
be impeded if consultants and managers \textbf{see privacy as a security
problem.} This is how manager M4 explained what their responsibility is
related to privacy:
\quoting{M4}{`And from my side, it's about keeping the system
secure in terms of proper security authorizations. And of course, on the
technical side, Microsoft already gives a certificate. }.

Furthermore, consultants and managers sometimes have
\textbf{no experience with PbD and DM in
the ERP system}. Thus, the ability to implement
these privacy techniques can be low:
\quoting{M2}{My experience is that I don't
have much experience.} However, \textbf{privacy out of the box} makes the
consultants' and managers' behavior
``\emph{easier}''. This way, consultants and managers perceive that they do not have to
implement privacy concepts themselves and think about whether the system
is GDPR-compliant.
\quoting{M1}{Like I said, the ERP system is
already compliant. So you don't have to be mindful of
the GDPR part}.

Consultants and managers are triggered to perform certain behaviors.
\textbf{Customer involvement} and \textbf{customer requirements} can be important triggers because the customer gives input on what data should be in the system together
with the consultant:
\quoting{M2}{Requirements are collected from the customer. The
customer must say that these are our requirements in the field of
privacy. We do not want everyone to be able to view the data of others.
If they're going to use the HR module within the ERP system, all your
details are listed there, so their address and bank details. Sometimes
the salary is also paid via the ERP system. Then the customer will
undoubtedly have requirements regarding privacy}.

As mentioned earlier, an \textbf{organization's privacy
policy} motivates consultants and managers, but it can also be a trigger
for consultants and managers to think about how
to manage sensitive information in the system. 

\textbf{Improvements of behavior.}
A variety of recommendations given by managers and consultants can help
to improve this behavior. One of the recommendations is to appoint a
\textbf{data protection officer (DPO)} for a project. This
recommendation makes the consultants' and
managers' behavior ``\emph{easier}'':
\quoting{M1}{The first thing I would say is [...] do you have a data protection officer or data controller as part
of your projects? That is the most important or the first question}. 
Another recommendation is establishing a \textbf{GDPR checklist} to make
the consultants' and managers' work more accessible. This checklist ensures that every release within a
project complies with GDPR, which makes thinking about GDPR during the
project much more manageable.
Having \textbf{general guidelines} can be sufficient for the managers or
consultants to increase their ability:
\quoting{M1}{And if I have the general guidelines on how this
data should be handled, I think that is sufficient for
us}.

The ability of consultants or managers can also be improved or made
more accessible by providing privacy modules in the ERP system. M6
explained that implementing privacy modules manually takes a lot of time and
effort: \quoting{M6}{And it would be good if in the future, from Salesforce or SAP, they would
already take that into account. So look at what is the law and regulations? What
is privacy? And then offer modules so you don’t have to create all that by hand}.

\textbf{Raising privacy awareness} is also a recommendation to improve
the ability of the consultants and managers. This will increase the
knowledge, and as a result, the consultants and managers will be able to
implement privacy techniques more easily and without problems. 

The last recommendation that can improve the ability of consultants and managers is to \textbf{involve everyone} during the project. For example, M2 explained that it is wise to bring developers, consultants, and everyone involved in a project together and talk to them to solve privacy problems:
\quoting{M2}{Another thing that is also important is that consultants and developers sit together. Then we can discuss what the challenges are in this area and how we can solve them, what we can and cannot do during the implementation}.

%% file: sections/discussion.tex
\section{Discussion}\label{discussion}

Developers, consultants, and managers behave almost the same way when it
comes to privacy in ERP systems. Ultimately, their behaviors can be improved following the given recommendations. We note that consultants and managers provided more recommendations to improve behavior compared to developers. For example, managers suggested hiring a data protection officer for a project and implementing privacy modules in the ERP system. 

It is known that PbD is not a very well-known concept for managers~\cite{vanrest2014designing}. The literature also shows that DM is a challenging concept for developers because they are not attuned to the privacy risks posed by the collected data and users' concerns about their privacy~\cite{schiffner2018towards}. Our research shows that these observations are still relevant when considering how managers and developers behave regarding privacy techniques in the ERP system. The literature proposes a DM methodology for software systems~\cite{senarath2019data}. However, our findings show that neither developers nor consultants adhere to this methodology. Moreover, \emph{Privacy Impact Assessment} (PIA) provides clear privacy objectives and specifies a means to achieve them. It is also known as a ``\emph{milestone towards privacy-by-design}''~\cite{edps2018}. However, PIA was not mentioned by our interviewees as a solution for implementing PbD.

%% file: sections/related_work.tex
\section{Related Work}\label{sec:relwork}

While security and privacy perceptions and behaviors of developers, security professionals, and other stakeholders are an active research area~\cite{kudriavtseva2025my,wee2024have,binkhorst2022security,van2024department,boteju2023sok,nurgalieva2023narrative}, to the best of our knowledge, no research has been done on identifying consultants' and developers' behavioral models related to ERP system privacy. 
 
\paragraph{Developers' privacy perceptions.}
Developers often believe that anonymizing data is more effective than privacy laws and practices in minimizing privacy concerns, and are often more prepared, compared to users, to give up privacy in return for better system functionality~\cite{sheth2014us}. They also might not always agree on what constitutes personal information~\cite{ma2024programmer} and overestimate their privacy inclinations and collect more data than their privacy attitudes might indicate~\cite{vanderlinden2021data}. 
Furthermore, it has been previously reported that developers frequently view privacy from a data security perspective and focus on technical solutions against data-related threats (access control, encryption, and anonymization)~\cite{hadar2018privacy,iwaya2023privacy,peixoto2023perspective,prybylo2024evaluating}. 
This is consistent with our findings.
Studies~\cite{hadar2018privacy,iwaya2023privacy} also found that the developers' work environment, namely the company’s privacy culture, influences their privacy perceptions and beliefs. Thus, organizations can affect developers' behavior regarding privacy engineering by creating and facilitating a privacy culture~\cite{iwaya2022organisational}.  Privacy regulations like GDPR positively affected developers' behaviors and organizations' privacy cultures~\cite{iwaya2023privacy}. This is consistent with our findings.

\paragraph{Other stakeholders.}
To the best of our knowledge, there is limited literature research about the privacy perceptions of managers or consultants. Henderson et al.~\cite{henderson1999personal} focuses on the actions an information system (IS) manager needs to take if there is a privacy concern within a system, but not on how the managers perceive privacy afforded by an IS system. Abomhara et al.~\cite{abomhara2024enhancing} studied the perspectives and attitudes of stakeholders engaged in national identification systems, reporting that there was a negative correlation between knowledge about PbD techniques and attitude towards it, explained by the perceived complexity. Closest to our work, Dalela et al.~\cite{dalela2021mixed} conducted a study of security and privacy practices in Danish companies, which included both managers and developers as participants. They found that managers and developers alike require better awareness about privacy practices and that GDPR was an important push towards improved data collection and protection methods.

%% file: sections/conclusions.tex
\section{Conclusions}\label{sec:conclusions}

In this study, we investigated the behavior of developers and consultants regarding GDPR compliance of ERP systems, and specifically their knowledge and perceptions regarding privacy-by-design and data minimization in the ERP system. We conducted semi-structured interviews and applied thematic analysis to discern developers' and consultants' perceptions and behavior following Fogg's Behavioral Model. We found that, while the enactment of GDPR had a major influence on privacy practices, there are still knowledge gaps and limited ability, motivation, and triggers to follow GDPR in ERP systems.